\documentclass[a4paper, 10pt, oneside, onecolumn, notitlepage, final]{article}
\usepackage[text={160mm, 240mm}, left=25mm, vmarginratio=1:1]{geometry}
\usepackage[T1]{fontenc}
\usepackage[utf8]{inputenc}
\usepackage{graphicx}
\usepackage{amssymb, amsfonts, amsmath, amsthm, mathtools, mathrsfs}
\usepackage{upgreek}
\usepackage[usenames, dvipsnames]{xcolor}
\usepackage{chemarrow}
\usepackage[colorlinks, linkcolor=blue, anchorcolor=red, citecolor=blue]{hyperref}
\usepackage[backend=biber, sorting=none, maxbibnames=3, minbibnames=3, style=numeric-comp]{biblatex}
\usepackage{braket} 
\usepackage{authblk} 
\usepackage{array}
\usepackage{bigstrut}
\usepackage{abstract}
\usepackage{titlesec}
\usepackage{csquotes}
\usepackage{lmodern}
\usepackage[english]{babel}

\titleformat{\section}{\large\bfseries\centering}{\thesection}{0.5 em}{}
\titleformat{\subsection}{\normalsize\bfseries\centering}{\thesubsection}{0.5 em}{}

\begin{document}

\title{\large\bfseries Tensor-Network Approaches to Counting Statistics for the Current in a Boundary-Driven Diffusive System}

\author[ ]{\normalsize Jiayin Gu\thanks{\texttt{gujiayin@pku.edu.cn}}}
\author[ ]{\normalsize Fan Zhang\thanks{\texttt{van314159@pku.edu.cn}}}
\affil[ ]{\normalsize School of Physics, Peking University, Beijing 100871, China}

\date{\vspace{-1cm}}
\maketitle

\begin{abstract}
We apply tensor networks to counting statistics for the stochastic particle transport in an out-of-equilibrium diffusive system. This system is composed of a one-dimensional channel in contact with two particle reservoirs at the ends. Two tensor-network algorithms, namely, Density Matrix Renormalization Group (DMRG) and Time Evolving Block Decimation (TEBD), are respectively implemented. The cumulant generating function for the current is numerically calculated and then compared with the analytical solution. Excellent agreement is found, manifesting the validity of these approaches in such an application. Moreover, the fluctuation theorem for the current is shown to hold.
\end{abstract}

\section{Introduction}

\par Away from equilibrium, open systems composed of atoms and molecules manifest macroscopic currents dissipating energy and producing thermodynamic entropy~\cite{Prigogine_1967, Nicolis_RepProgPhys_1979, deGroot_1984, Callen_1985}. These currents are intrinsically fluctuating with the probabilistic nature fully characterized by their large deviation properties~\cite{Ellis_2006, Vulpiani_2014, Touchette_PhysRep_2009}. The theoretical progress in the last three decades reveals that, for nonequilibrium systems in steady state, there exists a strikingly simple and general relation for the large deviation properties of the fluctuating currents. This relation is nowadays known as the fluctuation theorem~\cite{Evans_PhysRevLett_1993, Gallavotti_PhysRevLett_1996, Kurchan_JPhysA_1998, Lebowitz_JStatPhys_1999, Gaspard_JChemPhys_2004, Andrieux_JChemPhys_2004, Andrieux_JStatMech_2006, Derrida_JStatMech_2007}. At the mesoscopic level of description, the evolution of the coarse-grained state of the concerned system can be modeled as a Markov jump process. Accordingly, the probability distribution for the system states obeys a master equation. To evaluate the large deviation function for the current fluctuations, a direct method is to numerically simulate the Markov jump process with the Gillespie algorithm~\cite{Gillespie_JComputPhys_1976, Gillespie_JPhysChem_1977}, which generates a sample of random trajectories. Then, we perform counting statistics for the integrated current, and the probability distribution of the current can be finally constructed from the statistics~\cite{Touchette_arXiv_2012}. However, such direct sampling method encounters problem as it can not accurately estimate probability of rare events or large deviations given limited number of data. An alternative method is to first calculate the cumulant generating function for the currents, and then obtain the large deviation function through the Legendre-Fenchel transform. The cumulant generating function turns out to be the leading eigenvalue of the tilted generator of master equation including counting parameters~\cite{Touchette_arXiv_2012}. Thus, it has transformed into an eigenvalue problem.

\par For a spatially extended system, a coarse-grained description at the mesoscopic level requires partitioning the system in space. This leads to the characterization of the states with many random variables. Accordingly, the dimension of the tilted generator is very large, growing exponentially with the number of these variables. The same issue arises in quantum many-body systems for their underlying Hilbert space. Over the past few decades, tensor networks~\cite{Orus_NatRevPhys_2019, Orus_AnnPhys_2014, Verstraete_AdvPhys_2008, Schollwock_RevModPhys_2005, Schollwock_AnnPhys_2011, Cirac_RevModPhys_2021, Montangero_2018, Ran_2020} have emerged as one of the most powerful tools for handling such numerical complexity. Although originally developed in the context of condensed matter physics, tensor networks are presently finding applications in stochastic dynamics~\cite{Hieida_JPhysSocJpn_1998, Carlon_EurPhysJB_1999, Degenhard_MultiscaleModelSimul_2004, Helms_PhysRevE_2019, Helms_PhysRevLett_2020, Nagy_JStatPhys_2002, Temme_PhysRevLett_2010, Johnson_PhysRevE_2010, Gorissen_PhysRevE_2009, Gorissen_PhysRevE_2012, Gorissen_JPhysA_2011, Johnson_PhysRevLett_2015, Banuls_PhysRevLett_2019, Causer_arXiv_2022, Strand_arXiv_2022a, Strand_arXiv_2022b}. Indeed, there is a close analogy between the classical stochastic systems in nonequilibrium steady state and the quantum many-body systems in groud states. In this regard, two tensor-network approaches stand out, namely, Density Matrix Renormalization Group (DMRG) \cite{White_PhysRevLett_1992, White_PhysRevB_1993, Schollwock_RevModPhys_2005, Schollwock_AnnPhys_2011} and Time Evolving Block Decimation (TEBD)~\cite{Vidal_PhysRevLett_2003, Vidal_PhysRevLett_2004}. For quantum many-body systems, the former is a variational approach that enables the systematic search for ground states, while the latter an approach that evolves the states. If implemented in imaginary time, the latter can also be used to find the ground states. Correspondingly, for classical stochastic systems, these two approaches are naturally well suited for obtaining the nonequilibrium steady states.

\par Here, our purpose is to provide a pedagogical demonstration of how DMRG and TEBD are used to calculate the cumulant generating function for the current in a boundary-driven diffusive system. The rest of this paper is structured as follows. The stochastic description for the diffusive process is introduced in Section~\ref{sec_diffusive_system} where we discretize the system in space and then establish the master equation. The counting parameter for the particle transitions is included in the master equation, and consequently the cumulant generating function for the current is analytically obtained as the leading eigenvalue. Section~\ref{sec_DMRG} is devoted to the DMRG approach to counting statistics. The probability distribution is represented as a matrix product state (MPS). The generator for the master equation is formulated in terms of the local creation, annihilation, and particle number operators, and is then constructed as a matrix product operator (MPO). In Section~\ref{sec_TEBD}, the TEBD approach to counting statistics is concerned. The small-time evolution operator for the tilted master equation is approximately decomposed into a ordered product of local ones. The numerical details and results are presented in Section~\ref{sec_results}, followed by the concluding remarks given in Section~\ref{sec_conclusion}. The cumulant generating function is formally derived in Appendix~\ref{app_CGF}.

\section{Boundary-Driven Diffusive System}\label{sec_diffusive_system}

\begin{figure}
\begin{center}
\begin{minipage}[t]{0.6\hsize}
\resizebox{1.0\hsize}{!}{\includegraphics{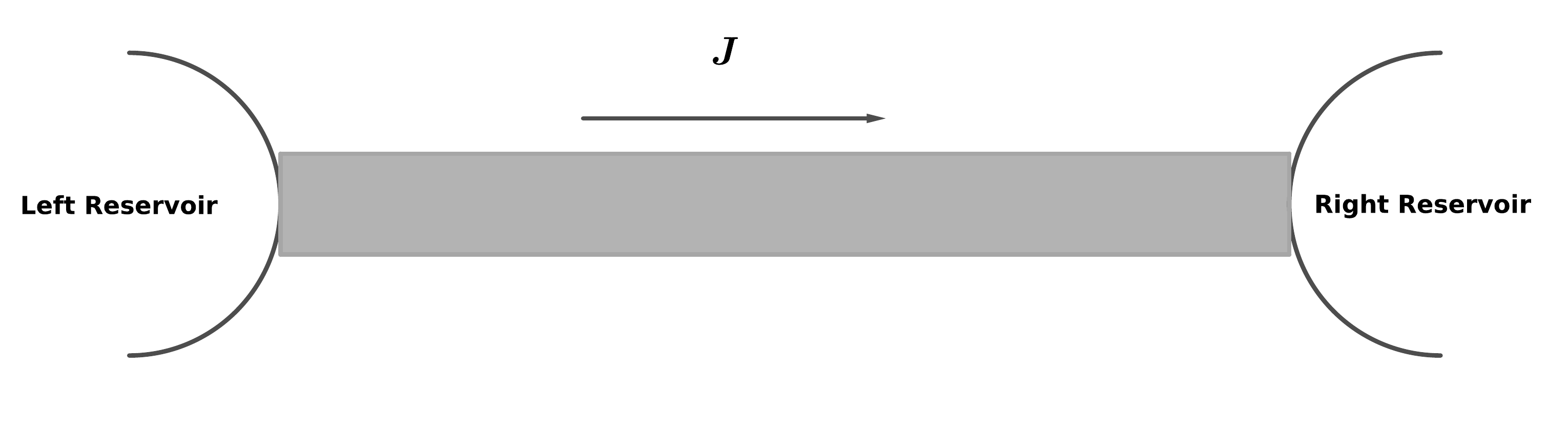}}
\end{minipage}
\end{center}
\caption{Schematic representation of a transport channel in contact with two reservoirs.}
\label{fig_channel}
\end{figure}

\par The system we consider here is modeled as a one-dimensional channel in contact with two reservoirs at the ends, as illustrated in Figure~\ref{fig_channel}. Distributed inside the channel are mobile particles with their densities expressed as a function of position, $n(x)$. At the boundaries, because the reservoirs are supposed to be arbitrarily large, the densities of particles are constant in time with $n_{\rm L}$ for the left reservoir and $n_{\rm R}$ for the right reservoir. In the case that $n_{\rm L}\neq n_{\rm R}$, the system is driven out of equilibrium and a directional current flows from one reservoir to the other. The mobile particles undergo the Brownian motion~\cite{Einstein_1956, Uhlenbeck_PhysRev_1930, Wang_RevModPhys_1945, Mazo_2002}, so that the time evolution of the densities is given by the diffusion equation
\begin{align}
\frac{\partial n(x,t)}{\partial t}=D\frac{\partial^2n(x,t)}{\partial x^2} \text{,}
\end{align}
where $D$ is the diffusion coefficient determined phenomenally by direct observation of the random walk of a tagged Brownian particle according to the formula
\begin{align}
D=\lim_{t\to\infty}\frac{1}{2t}\left\langle(x_t-x_0)^2\right\rangle \text{.}
\end{align}
Here, $x_0$ (respectively, $x_t$) is the position at the initial time (respectively, time $t$), and $\langle\cdot\rangle$ denotes the average taken over the trajectories.

\subsection{Stochastic Description}

\par At the mesoscopic level, the time evolution of the distribution of mobile particles in the channel can be described as a Markov jump process, which is formulated in terms of a master equation (see Ref.~\cite{Gaspard_NewJPhys_2005} and references therein). The fluctuations down to the mesoscopic scale can be fully characterized by such a stochastic description. For this purpose, the channel is spatially discretized into $L$ cells, each of width $\Delta x$ and containing some number of mobile particles. These cells are labeled with indices $i=1,2,\cdots,L$. For notational convenience, the indices $i=0$ and $i=L+1$ are respectively used to refer to the cells for left and right reservoirs. The number of particles in both cells for reservoirs are maintained constant in time, and are given by $N_0\equiv\bar{N}_{\rm L}$ and $N_{L+1}\equiv\bar{N}_{\rm R}$. The system state is specified by the numbers ${\bf N}=\{N_i\}_{i=1}^L$ of particles in the cells. These numbers change in time due to the random transitions according to the kinetic network
\begin{equation}
\resizebox{1.0\hsize}{!}{$
\begin{array}{ccccccccccccc}
\bar{N}_{\rm L} & \autorightleftharpoons{$\scriptstyle W_0^{(+)}$}{$\scriptstyle W_0^{(-)}$} & N_1 & \autorightleftharpoons{$\scriptstyle W_1^{(+)}$}{$\scriptstyle W_1^{(-)}$}& N_2 & \autorightleftharpoons{$\scriptstyle W_2^{(+)}$}{$\scriptstyle W_2^{(-)}$} & \cdots & \autorightleftharpoons{$\scriptstyle W_{L-2}^{(+)}$}{$\scriptstyle W_{L-2}^{(-)}$} & N_{L-1}& \autorightleftharpoons{$\scriptstyle W_{L-1}^{(+)}$}{$\scriptstyle W_{L-1}^{(-)}$} & N_L & \autorightleftharpoons{$\scriptstyle W_L^{(+)}$}{$\scriptstyle W_L^{(-)}$} & \bar{N}_{\rm R} \nonumber \text{,}
\end{array}$}
\end{equation}
with the transition rates linearly depending on the particle number in the departure cell, $W_i^{(+)}=kN_i$, $W_i^{(-)}=kN_{i+1}$.
Here, $k$ is the rate constant related to $D$ by $k=D/\Delta x^2$ which can be obtained through the continuum limit\footnote{The mean current is expressed as
\begin{align}
J=-k\langle N_{i+1}\rangle+k\langle N_i\rangle=-k\Omega\Delta x\frac{\partial n}{\partial x}=j\Sigma=-D\frac{\Omega}{\Delta x}\frac{\partial n}{\partial x} \text{,} \nonumber
\end{align}
where $\Omega$ and $\Sigma$ denote the volume and section area of a cell, respectively. From this expression, $k=D/\Delta x^2$ is immediately arrived.}.
The probability ${\cal P}({\bf N},t)$ that the cells contain the particle numbers ${\bf N}$ for time $t$ obeys the master equation 
\begin{align}
\frac{{\rm d}{\cal P}}{{\rm d}t}=\hat{L}{\cal P}= & k\bar{N}_{\rm L}{\cal P}(N_1-1,\cdots,t)-k\bar{N}_{\rm L}{\cal P}(N_1,\cdots,t) \nonumber \\
& +k(N_1+1){\cal P}(N_1+1,\cdots,t)-kN_1{\cal P}(N_1,\cdots,t) \nonumber \\
& \hspace{1cm}\vdots \nonumber \\
& +k(N_i+1){\cal P}(\cdots,N_i+1,N_{i+1}-1,\cdots,t)-kN_i{\cal P}(\cdots,N_i,N_{i+1},\cdots,t) \nonumber \\
& +k(N_{i+1}+1){\cal P}(\cdots,N_i-1,N_{i+1}+1,\cdots,t)-kN_{i+1}{\cal P}(\cdots,N_i,N_{i+1},\cdots,t) \nonumber \\
& \hspace{1cm}\vdots \nonumber \\
& +k(N_L+1){\cal P}(\cdots,N_L+1,t)-kN_L{\cal P}(\cdots,N_L,t) \nonumber \\
& +k\bar{N}_{\rm R}{\cal P}(\cdots,N_L-1,t)-k\bar{N}_{\rm R}{\cal P}(\cdots,N_L,t) \text{,} \label{eq_master_equation}
\end{align}
where $\hat{L}$ denotes the generator for the time evolution. The stationary solution of the above master equation is given by a product of Poisson distributions,
\begin{align}
{\cal P}_{\rm st}(N_1,\cdots,N_L)=\prod_{i=1}^L\frac{\langle N_i\rangle_{\rm st}^{N_i}}{N_i!}{\rm e}^{{-\langle N_i\rangle}_{\rm st}} \text{,} \label{eq_joint_P}
\end{align}
with the mean number of particles corresponding to a linear profile of concentration,
\begin{align}
\langle N_i\rangle_{\rm st}=\bar{N}_{\rm L}+\frac{\bar{N}_{\rm R}-\bar{N}_{\rm L}}{L+1}i \text{.} \label{eq_mean}
\end{align}

\subsection{Counting Statistics}

\par To investigate the current fluctuations for the particle transport, we now perform counting statistics. For this purpose, the master equation~(\ref{eq_master_equation}) is modified to include a counting parameter $\lambda$ for particle transitions between the left-reservoir cell and the first one for the discretized channel, yielding
\begin{align}
\frac{{\rm d}F}{{\rm d}t}=\hat{L}_{\lambda}F= & k\bar{N}_{\rm L}{\rm e}^{-\lambda}F(N_1-1,\cdots,t)-k\bar{N}_{\rm L}F(N_1,\cdots,t) \nonumber \\
& +k(N_1+1){\rm e}^{+\lambda}F(N_1+1,\cdots,t)-kN_1F(N_1,\cdots,t) \nonumber \\
& \hspace{1cm}\vdots \nonumber \\
& +k(N_i+1)F(\cdots,N_i+1,N_{i+1}-1,\cdots,t)-kN_iF(\cdots,N_i,N_{i+1},\cdots,t) \nonumber \\
& +k(N_{i+1}+1)F(\cdots,N_i-1,N_{i+1}+1,\cdots,t)-kN_{i+1}F(\cdots,N_i,N_{i+1},\cdots,t) \nonumber \\
& \hspace{1cm}\vdots \nonumber \\
& +k(N_L+1)F(\cdots,N_L+1,t)-kN_LF(\cdots,N_L,t) \nonumber \\
& +k\bar{N}_{\rm R}F(\cdots,N_L-1,t)-k\bar{N}_{\rm R}F(\cdots,N_L,t) \text{.} \label{eq_tilted_master_equation}
\end{align}
The eigensolution of this tilted master equation has the form
\begin{align}
F(N_1,\cdots,N_L,t)={\rm e}^{-Q(\lambda)t}\prod_{i=1}^L\frac{C_i^{N_i}}{N_i!} \text{,} \label{eq_joint_F}
\end{align}
where $\{C_i\}$ are constants depending on $\lambda$, and $Q(\lambda)$ is the cumulant generating function. Substituting this solution into the tilted master equation and after some algebraic manipulations, we obtain
\begin{align}
Q(\lambda)=\frac{k\bar{N}_{\rm L}}{L+1}\left(1-{\rm e}^{-\lambda}\right) +\frac{k\bar{N}_{\rm R}}{L+1}\left(1-{\rm e}^{\lambda}\right) \text{.} \label{eq_CGF}
\end{align}
If we define $A=\ln\left(\bar{N}_{\rm L}/\bar{N}_{\rm R}\right)$ as the affinity~\cite{Donder_1936} for driving the system out of equilibrium, Then, the cumulant generating function~(\ref{eq_CGF}) satisfies the Gallavotti-Cohen symmetry~\cite{Gallavotti_PhysRevLett_1995, Gallavotti_JStatPhys_1995, Lebowitz_JStatPhys_1999}
\begin{align}
Q(\lambda)=Q(A-\lambda) \text{,} \label{eq_symmetry}
\end{align}
which is also known under the name fluctuation theorem. The statistical cumulants can be obtained by taking successive derivatives with respect to the counting parameter $\lambda$. For example, the mean current and its diffusivity are given by
\begin{align}
& J=\left.\frac{\partial Q}{\partial\lambda}\right\vert_{\lambda=0}=\frac{k\left(\bar{N}_{\rm L}-\bar{N}_{\rm R}\right)}{L+1} \text{,} \\
& D=-\frac{1}{2}\left.\frac{\partial^2 Q}{\partial\lambda^2}\right\vert_{\lambda=0}=\frac{k\left(\bar{N}_{\rm L}+\bar{N}_{\rm R}\right)}{2(L+1)} \text{.}
\end{align}
For nonequilibrium systems described by master equation, the affinity can be systematically obtained through Schnakenberg graph analysis~\cite{Schnakenberg_RevModPhys_1976}. More details about the counting statistics for this diffusive process can be found in Ref.~\cite{Andrieux_JStatMech_2006}. A formal derivation of the cumulant generating function~(\ref{eq_CGF}) is provided in Appendix~\ref{app_CGF}.

\par From the numerical perspective, the determination of cumulant generating function requires to calculate the leading eigenvalue of the tilted generator $\hat{L}_{\lambda}$ defined in Eq.~(\ref{eq_tilted_master_equation}). In principle, this can be done with various numerical routines for diagonalizing a matrix. However, this is often very difficult in practice. Since the dimension of the tilted generator grows exponentially with the number of the discretized cells. This motivates the need of tensor networks to tackle such a problem.

\section{DMRG Approach to Counting Statistics}\label{sec_DMRG}

\par In this section, we present in detail the DMRG procedure to perform counting statistics for the stochastic process in the preceding section. The key differences from the standard usage for quantum lattice systems are highlighted.

\subsection{Notations}

\par First, it is helpful to introduce the quantum-mechanical notation for subsequent description convenience. We use $\ket{N_i}$ for the state that the $i$-th cell contains $N_i$ particle and $\ket{\bf N}$ for the system state specified by the numbers ${\bf N}=\{N_i\}_{i=1}^L$. Any distribution $F({\bf N})$ is written as
\begin{align}
\ket{F({\bf N})}=\sum_{\bf N}F(\bf N)\ket{\bf N} \text{.}
\end{align}
Accordingly, $\bra{N_i}$, $\bra{\bf N}$, and $\bra{F^{\dagger}(\bf N)}$ are used to denote their adjoints. Following the conventions in quantum mechanics, 2-norm and orthogonality are assumed,
\begin{align}
& \braket{N'|N}=\delta_{N',N} \text{,} \\
& \braket{{\bf N}'|{\bf N}}=\delta_{{\bf N}',{\bf N}} \text{,} \\
& \braket{F^{\dagger}({\bf N})|F({\bf N})}=\sum_{\bf N}F^{\dagger}({\bf N})F({\bf N})=1 \text{.}
\end{align}
Now, an issue arises that the probability distribution is of 1-norm, $\sum_{\bf N}{\cal P}({\bf N})=1$. We can introduce the 2-norm probability distribution,
\begin{align}
\tilde{\cal P}({\bf N})\equiv\frac{{\cal P}({\bf N})}{\sqrt{\sum_{\bf N}{\cal P}({\bf N}){\cal P}({\bf N})}} \text{,} \label{eq_2norm_prob}
\end{align}
so that $\braket{\tilde{\cal P}^{\dagger}({\bf N})|\tilde{\cal P}({\bf N})}=1$. The 1-norm probability distribution can be recovered,
\begin{align}
{\cal P}({\bf N})=\frac{\tilde{\cal P}({\bf N})}{\sum_{\bf N}\tilde{\cal P}({\bf N})} \text{.} \label{eq_1norm_prob}
\end{align}
There is a one-to-one correspondence between these two kinds of probability distributions. In practical implementation in code, we always use the 2-norm version, since many tensor-network techniques already developed for quantum-mechanical systems can be utilized.

\subsection{MPS and MPO}

\begin{figure}
\begin{center}
\begin{minipage}[t]{0.45\hsize}
\resizebox{1.0\hsize}{!}{\includegraphics{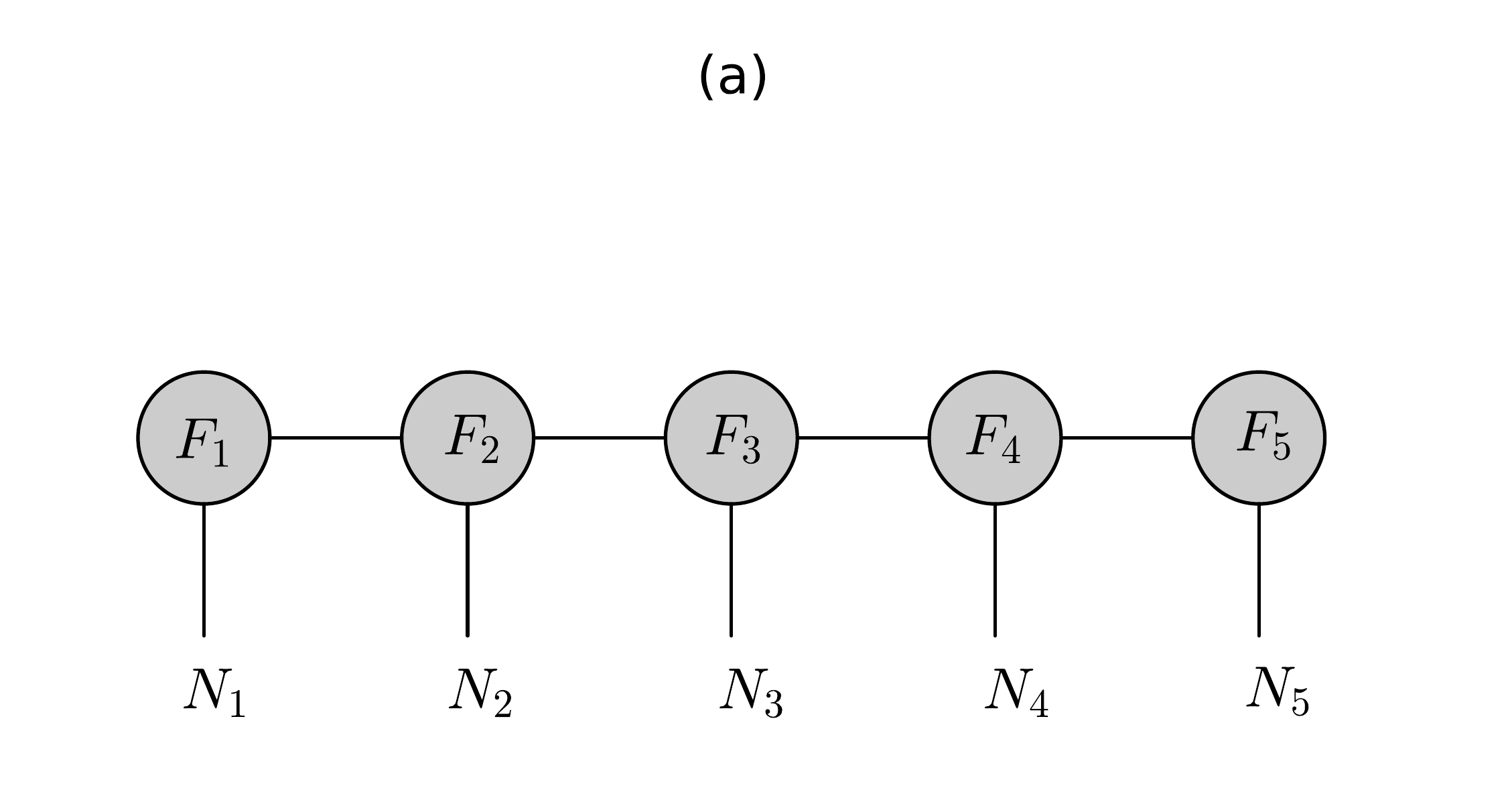}}
\end{minipage}
\begin{minipage}[t]{0.45\hsize}
\resizebox{1.0\hsize}{!}{\includegraphics{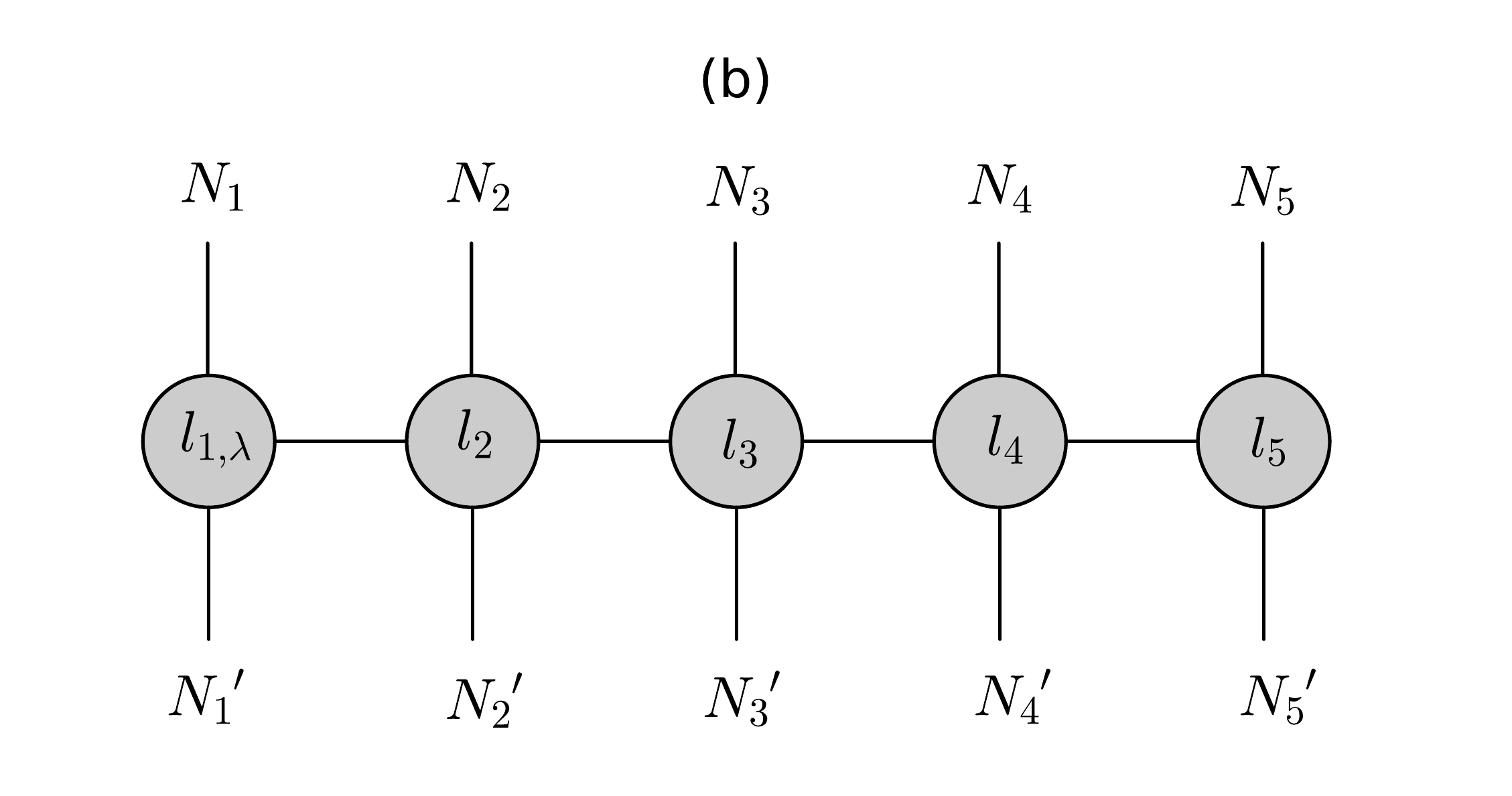}}
\end{minipage}
\end{center}
\caption{Graphical representation of (a) an MPS and (b) an MPO with the counting parameter $\lambda$ included in the first tensor.}
\label{fig_MPSMPO}
\end{figure}

\par In quantum many-body systems, the presence of the celebrated area laws~\cite{Eisert_RevModPhys_2010} for the entanglement entropy enables an efficient numerical representation of the ground state as an MPS~\cite{PerezGarcia_arXiv_2007} (see Figure~\ref{fig_MPSMPO}a for illustration). Here, MPS can also be used to represent the stationary solution for classical stochastic systems in nonequilibrium steady states. Moreover, we notice that the feature of joint product in solutions~(\ref{eq_joint_P}) and~(\ref{eq_joint_F}) imply that the bond dimensions of their MPS representations are both one. This is very remarkable that the computer memory used to store the stationary solution can be very low, scaling linearly with the number of cells. The physical indices denote the numbers $\{N_i\}$.

\par Next, we show how to construct an MPO (see Figure~\ref{fig_MPSMPO}b) for the tilted generator $\hat{L}_{\lambda}$. This constitutes the most crucial step in applying the DMRG approach to counting statistics for the stochastic process. We adopt the formalism similar to the second quantization. The particle transitions between two neighboring cells can be associated with an annihilation operator acting on one cell and an creation operator acting the other. Following this reasoning, we introduce the local annihilation operator $a_i^-$ and creation operator $a_i^+$, which are respectively defined by
\begin{align}
& \braket{N_i|a_i^-|N_i'}=kN_i'\delta_{N_i,,N_i'-1} \text{,} \label{eq_a_minus} \\
& \braket{N_i|a_i^+|N_i'}=\delta_{N_i,N_i'+1} \text{.} \label{eq_a_plus}
\end{align}
It should be noted that the operator definitions by Eqs.~(\ref{eq_a_minus})-(\ref{eq_a_plus}) only apply to the cells for the channel, $1\le i\le L$. For the cells corresponding to the reservoirs, $i=0,L+1$, these operators should be defined differently,
\begin{align}
& a_0^-=k\bar{N}_{\rm L} \text{,} \label{eq_a_boundaries_1} \\
& a_0^+=a_{L+1}^+=1 \text{,} \label{eq_a_boundaries_2} \\
& a_{L+1}^-=k\bar{N}_{\rm R} \text{.} \label{eq_a_boundaries_3}
\end{align}
Because the particle numbers in the reservoir cells are kept constant. To account for the probability loss in the tilted master equation, we should still introduce another operator $a_i$ defined by
\begin{align}
\braket{N_i|a_i|N_i'}=kN_i'\delta_{N_i,N_i'} \text{,} \label{eq_a}
\end{align}
which can be intuitively termed as particle number operator. This definition applies to any cell. In this way, the tilted generator is expressed as
\begin{align}
\hat{L}_{\lambda}=a_0^-\otimes a_1^+{\rm e}^{-\lambda}+a_0^+\otimes a_1^-{\rm e}^{+\lambda}+\sum_{i=1}^L\left(a_i^-\otimes a_{i+1}^++a_i^+\otimes a_{i+1}^-\right)-\sum_{i=0}^L\left(a_i+a_{i+1}\right) \text{,} \label{eq_L_lambda}
\end{align}
where $\otimes$ denotes the tensor product and the counting parameter $\lambda$ is included for particle transitions between the left-reservoir cell and the first one in the discretized channel. This tilted generator is now split into two parts
\begin{align}
\hat{L}_{\lambda}=\hat{L}_{\lambda}^{(1)}+\hat{L}^{(2)} \text{,}
\end{align}
with each constructed as an MPO,
\begin{align}
\hat{L}_{\lambda}^{(1)} & =a_0^-\otimes a_1^+{\rm e}^{-\lambda}+a_0^+\otimes a_1^-{\rm e}^{+\lambda}+\sum_{i=1}^L\left(a_i^-\otimes a_{i+1}^++a_i^+\otimes a_{i+1}^-\right) \nonumber \\
& =\begin{pmatrix}
a_0^-{\rm e}^{-\lambda} & a_0^+ & 0 & 1
\end{pmatrix}\otimes\begin{pmatrix}
0 & 0 & a_1^+ & 0 \\
0 & 0 & a_1^-{\rm e}^{+\lambda} & 0 \\
0 & 0 & 1 & 0 \\
a_1^- & a_1^+ & 0 & 1 
\end{pmatrix}\otimes\cdots\otimes\begin{pmatrix}
0 & 0 & a_i^+ & 0 \\
0 & 0 & a_i^- & 0 \\
0 & 0 & 1 & 0 \\
a_i^- & a_i^+ & 0 & 1 
\end{pmatrix}\otimes\cdots\otimes\begin{pmatrix}
a_{L+1}^+ \\
a_{L+1}^- \\
1 \\
0
\end{pmatrix} \nonumber \\
& =\begin{pmatrix}
a_1^- & a_1^+ & a_1^+k\bar{N}_{\rm L}{\rm e}^{-\lambda}+a_1^-{\rm e}^{+\lambda} & 1
\end{pmatrix}\otimes\cdots\otimes\begin{pmatrix}
0 & 0 & a_i^+ & 0 \\
0 & 0 & a_i^- & 0 \\
0 & 0 & 1 & 0 \\
a_i^- & a_i^+ & 0 & 1 
\end{pmatrix}\otimes\cdots\otimes\begin{pmatrix}
a_L^+ \\
a_L^- \\
1 \\
a_L^-+a_L^+k\bar{N}_{\rm R}
\end{pmatrix}
\end{align}
and
\begin{align}
\hat{L}^{(2)} & =\sum_{i=0}^L\left(-a_i-a_{i+1}\right) \nonumber \\
& =\begin{pmatrix}
1 & -a_0 & 1
\end{pmatrix}\otimes\cdots\otimes\begin{pmatrix}
0 & -a_i & 0 \\
0 & 1 & 0 \\
1 & -a_i & 1
\end{pmatrix}\otimes\cdots\otimes\begin{pmatrix}
-a_{L+1} \\
1 \\
0
\end{pmatrix} \nonumber \\
& =\begin{pmatrix}
1 & -k\bar{N}_{\rm L}-2a_1 & 1
\end{pmatrix}\otimes\cdots\otimes\begin{pmatrix}
0 & -a_i & 0 \\
0 & 1 & 0 \\
1 & -a_i & 1
\end{pmatrix}\otimes\cdots\otimes\begin{pmatrix}
-a_L \\
1 \\
-a_L-k\bar{N}_{\rm R}
\end{pmatrix} \text{,}
\end{align}
where the boundary tensors are explicitly expressed according to Eqs.~(\ref{eq_a_boundaries_1})-(\ref{eq_a_boundaries_3}) and contracted with their neighboring ones. So, there are $L$ tensors in each MPO as expected. If we denotes
\begin{align}
\hat{L}_{\lambda}^{(1)}=l_{1,\lambda}^{(1)}\otimes\cdots\otimes l_i^{(1)}\otimes\cdots\otimes l_L^{(1)}
\end{align}
and
\begin{align}
\hat{L}^{(2)}=l_1^{(2)}\otimes\cdots\otimes l_i^{(2)}\otimes\cdots\otimes l_L^{(2)} \text{,}
\end{align}
then the tilted generator is constructed as an MPO,
\begin{align}
\hat{L}_{\lambda} = l_{1,\lambda}\otimes\cdots\otimes l_i\otimes\cdots\otimes l_L =\begin{pmatrix}
l_{1,\lambda}^{(1)} & l_1^{(2)}
\end{pmatrix}\otimes\cdots\otimes\begin{pmatrix}
l_i^{(1)} & 0 \\
0 & l_i^{(2)}
\end{pmatrix}\otimes\cdots\otimes\begin{pmatrix}
l_L^{(1)} \\
l_L^{(2)}
\end{pmatrix} \text{,}
\end{align}
whose bond dimension is 7. It should be noticed that the tilted generator $\hat{L}_{\lambda}$ is intrinsically non-symmetric or non-Hermitian, implying that (i) the right eigenvectors are not identical to the corresponding left eigenvectors; and (ii) the right (left) eigenvectors are not mutually orthogonal. This is the basic difference from the case of quantum Hamiltonian.

\subsection{Optimization}

\begin{figure}
\begin{center}
\begin{minipage}[t]{0.7\hsize}
\resizebox{1.0\hsize}{!}{\includegraphics{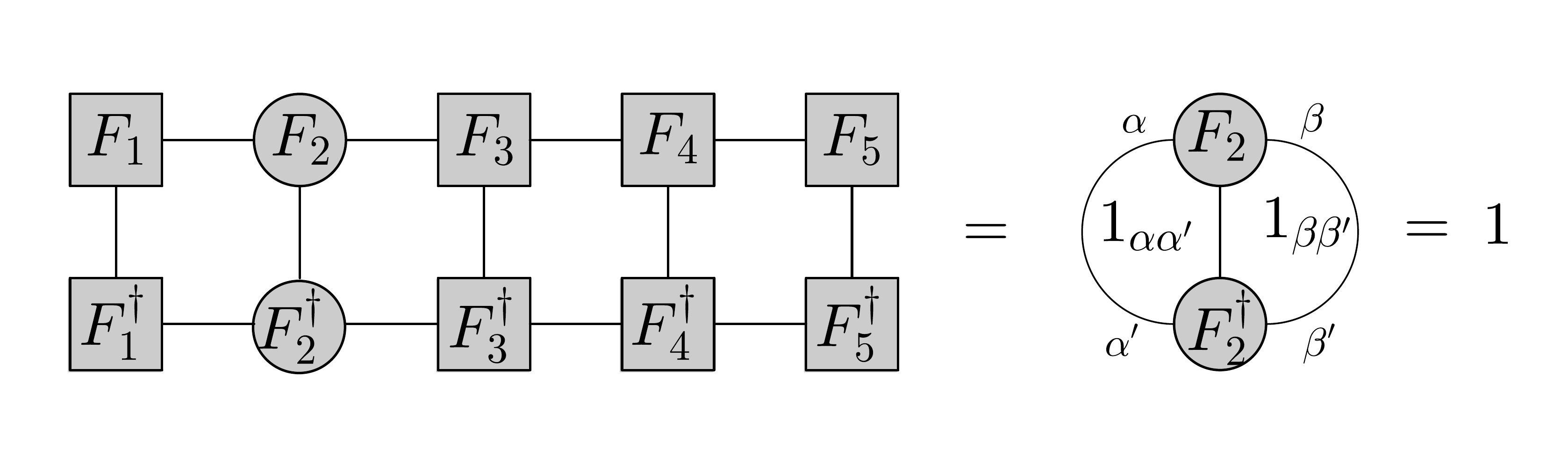}}
\end{minipage}
\end{center}
\caption{Graphical representation of the contraction between an normalized MPS and its adjoint. Both are in the canonical form with the centers lying at $F_2$ and $F_2^{\dagger}$, respectively.}
\label{fig_canonical_MPS}
\end{figure}

\par
\begin{figure}
\begin{center}
\begin{minipage}[t]{0.8\hsize}
\resizebox{1.0\hsize}{!}{\includegraphics{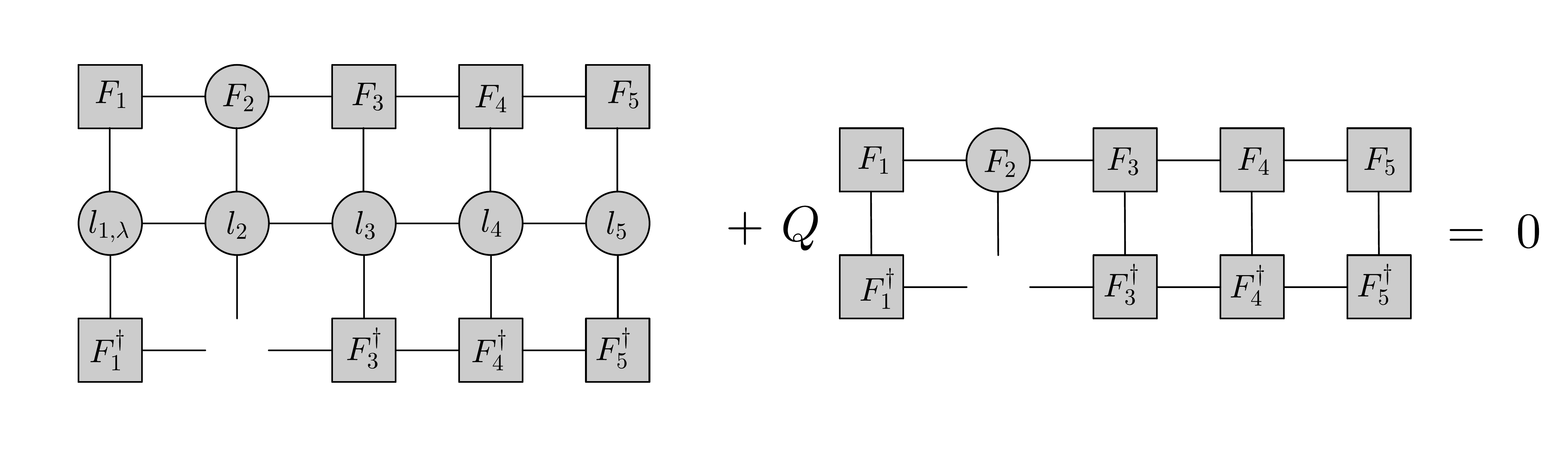}}
\end{minipage}
\end{center}
\caption{The graphical representation of the variational procedure~(\ref{eq_DMRG1}) with, e.g., $F_2$ being the canonical center.}
\label{fig_DMRG1}
\end{figure}

\begin{figure}
\begin{center}
\begin{minipage}[t]{0.8\hsize}
\resizebox{1.0\hsize}{!}{\includegraphics{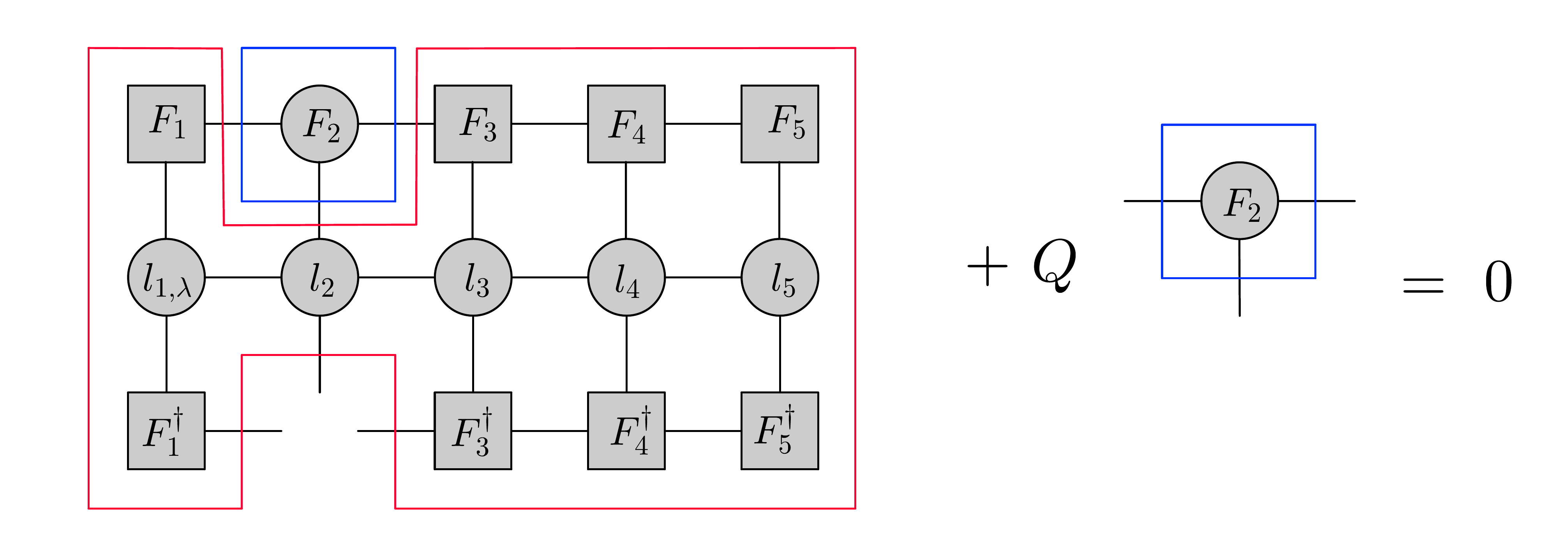}}
\end{minipage}
\end{center}
\caption{Graphical representation of the eigenvalue problem~(\ref{eq_DMRG2}).}
\label{fig_DMRG2}
\end{figure}

 The aim of DMRG approach is to variationally find the eigensolution from which the leading eigenvalue can be computed. This can be formulated as an optimization problem
\begin{align}
Q(\lambda)=-{\rm max}\left[\braket{F^{\dagger}({\bf N})|\hat{L}_{\lambda}|F({\bf N})}\right] \text{,} \label{eq_optimization_1}
\end{align}
under the condition $\braket{F^{\dagger}({\bf N})|F({\bf N})}=1$. The graphical representation of this normalization is shown in Figure~\ref{fig_canonical_MPS} where the canonical form of MPS is presented due to the gauge freedom. The advantage of this canonical form lies in the efficient contraction of an MPS with its adjoint; we only need to contract the canonical centers. The shift of the canonical center can be realized with the singular value decomposition. The problem~(\ref{eq_optimization_1}) can be further reformulated into a constrained optimization problem using a Lagrangian multiplier $-Q$,
\begin{align}
\braket{F^{\dagger}({\bf N})|\hat{L}_{\lambda}|F({\bf N})}+Q\braket{F^{\dagger}({\bf N})|F({\bf N})} \text{.} \label{eq_optimization_2}
\end{align}
We now adopt the single-tensor optimization procedure, where only one tensor $F_i^{\dagger}$ is varied at once while keeping all others fixed, so we have
\begin{align}
\frac{\partial}{\partial F_i^{\dagger}}\left[\braket{F^{\dagger}({\bf N})|\hat{L}_{\lambda}|F({\bf N})}+Q\braket{F^{\dagger}({\bf N})|F({\bf N})}\right]=0 \text{.} \label{eq_DMRG1}
\end{align}
Figure~\ref{fig_DMRG1} shows the graphical representation of the consequence after this variational procedure. Because two MPSs are both in canonical form with the centers at $F_i$ and $F_i^{\dagger}$, Eq.~(\ref{eq_DMRG1}) simplifies to
\begin{align}
\tilde{\hat{L}}_{\lambda}\cdot F_i+QF_i=0 \text{,} \label{eq_DMRG2}
\end{align}
where $\tilde{\hat{L}}_{\lambda}$ the operator indicated as the red part in Figure~\ref{fig_DMRG2}. This now represents a standard eigenvalue problem. The dimension of the operator $\tilde{\hat{L}}_{\lambda}$ is greatly reduced, making it possible to be handled with normal numerical routines.

\par We now need to find the right eigenvector $F_i$ of $\tilde{\hat{L}}_{\lambda}$ corresponding to the largest eigenvalue $-Q$. This can be achieved through the iterative method. Since ${\rm e}^{\tilde{\hat{L}}_{\lambda}}>0$, the Perron-Frobenius theorem applies, and the leading eigenvalue $-Q$ of $\tilde{\hat{L}}_{\lambda}$ corresponds to the real maximum eigenvalue ${\rm e}^{-Qt}$ of ${\rm e}^{\tilde{\hat{L}}_{\lambda}t}$ in magnitude (for some $t>0$). Consequently, the right eigenvector can be asymptotically evaluated as
\begin{align}
F_i\sim_{t\to\infty}{\rm e}^{\tilde{\hat{L}}_{\lambda}t}F_i(0) \text{,} \label{eq_F_i}
\end{align}
which is then normalized
\begin{align}
F_i\leftarrow\frac{F_i}{\sqrt{F^{\dagger}_i\cdot F_i}} \text{.}
\end{align}
Here, $F_i(0)$ is an arbitrary vector assumed to have the component of the desired eigenvector. The matrix exponential can be computed with many numerical routines~\cite{Moler_SIAMRev_1978}. Then, we have
\begin{align}
Q=-F_i^{\dagger}\cdot\tilde{\hat{L}}_{\lambda}\cdot F_i \text{.}
\end{align}
After completing the optimization on $F_i$, we move the canonical center to a neighboring one and continue such procedure. This operation is performed back and forth between the first and last tensors until the leading eigenvalue converges. With different values of $\lambda$, we can numerically construct the desired cumulant generating function $Q(\lambda)$.

\section{TEBD Approach to Counting Statistics}\label{sec_TEBD}

\par
\begin{figure}
\begin{center}
\begin{minipage}[t]{0.5\hsize}
\resizebox{1.0\hsize}{!}{\includegraphics{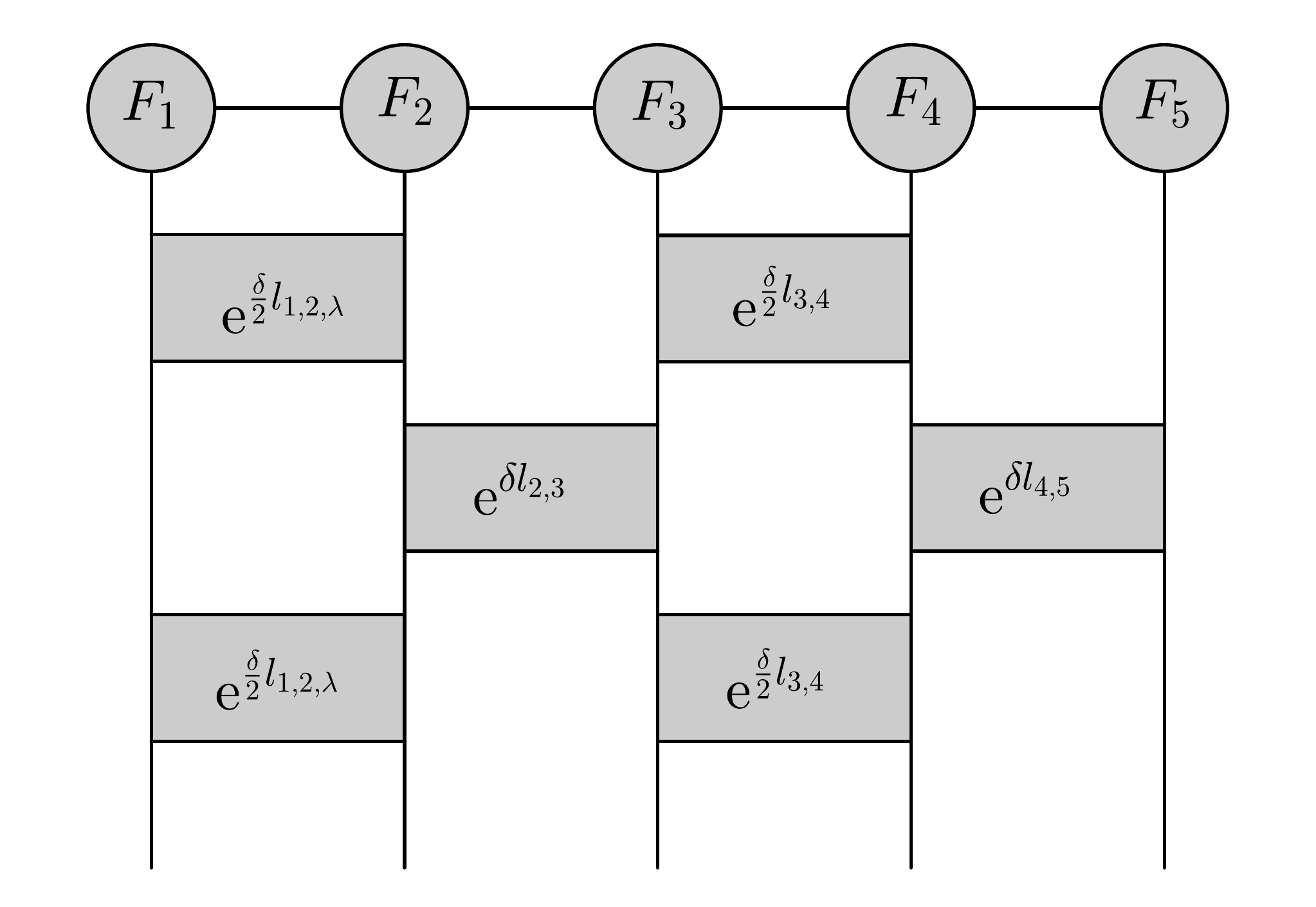}}
\end{minipage}
\end{center}
\caption{Graphical representation of the TEBD algorithm implemented with the second-order Trotter-Suzuki decomposition. The local evolution operators are alternatively applied to the MPS. The counting parameter $\lambda$ is included in $l_{1,2,\lambda}$.}
\label{fig_TEBD}
\end{figure}

 In this section, we present another tensor-network approach to counting statistics for the stochastic current. Now that we want to obtain the eigensolution of the tilted generator $\hat{L}_{\lambda}$ corresponding to the leading eigenvalue, can we achieve this with direct iteration? The answer to this question is yes. This can be done with TEBD which is an algorithm relying on Trotter-Suzuki decomposition~\cite{Suzuki_CommunMathPhys_1976}. The desired eigensolution is asymptotically evaluated as
\begin{align}
\ket{F({\bf N})}\sim_{t\to\infty}{\rm e}^{\hat{L}_{\lambda}t}\ket{F({\bf N},0)} \text{,}
\end{align}
where $\ket{F({\bf N},0)}$ is an arbitrary initial distribution. At this point, the dimension of the evolution operator is exponentially large. For small-time step $\delta$, however, ${\rm e}^{\hat{L}_{\lambda}\delta}$ can be decomposed into an ordered product of local ones of a much smaller dimension. Because the particles hop between neighboring cells, the tilted generator $\hat{L}_{\lambda}$ can be split into two parts,
\begin{align}
\hat{L}_{\lambda}=\hat{L}_{\rm odd}+\hat{L}_{\rm even} \text{,}
\end{align}
with
\begin{align}
& \hat{L}_{\rm odd}=l_{1,2,\lambda}+l_{3,4}+\cdots \text{,} \\
& \hat{L}_{\rm even}=l_{2,3}+l_{4,5}+\cdots \text{,}
\end{align}
where $l_{i,i+1}$ represents the local operator acting simultaneously on the $i$-th and $(i+1)$-th cells. The local operators in each part can be diagonalized efficiently and mutually commuting. It is noted here that the counting parameter is included in $l_{1,2,\lambda}$. The exact evolution operator can be decomposed to any order. Here, we give the second-order one,
\begin{align}
{\rm e}^{\delta\hat{L}_{\lambda}}\approx{\rm e}^{\frac{\delta}{2}\hat{L}_{\rm odd}}{\rm e}^{\delta\hat{L}_{\rm even}}{\rm e}^{\frac{\delta}{2}\hat{L}_{\rm odd}}+O(\delta^3) \text{,}
\end{align}
as it is commonly used. If the transport channel is discretized into $5$ cells, then the local operators are explicitly given by
\begin{align}
& l_{1,2,\lambda}=k\bar{N}_{\rm L}{\rm e}^{-\lambda}a_1^+\otimes{\sf I}_2+{\rm e}^{+\lambda}a_1^-\otimes{\sf I}_2+a_1^-\otimes a_2^++a_1^+\otimes a_2^--k\bar{N}_{\rm L}{\sf I}_1\otimes{\sf I}_2-2a_1\otimes{\sf I}_2-{\sf I}_1\otimes a_2 \text{,} \\
& l_{3,4}=a_3^-\otimes a_4^++a_3^+\otimes a_4^--a_3\otimes{\sf I}_4-{\sf I}_3\otimes a_4  \text{,} \\
& l_{2,3}=a_2^-\otimes a_3^++a_2^+\otimes a_3^--a_2\otimes{\sf I}_3-{\sf I}_2\otimes a_3 \text{,} \\
& l_{4,5}=a_4^-\otimes a_5^++a_4^+\otimes a_5^-+{\sf I}_4\otimes a_5^-+k\bar{N}_{\rm R}{\sf I}_4\otimes a_5^+-a_4\otimes{\sf I}_5-2{\sf I}_4\otimes a_5-k\bar{N}_{\rm R}{\sf I}_4\otimes{\sf I}_5 \text{,}
\end{align}
where ${\sf I}_i$ stands for the identity operator acting on the $i$-th cell. The local evolution operators are alternatively applied to the MPS (see Figure~\ref{fig_TEBD} for illustration). Once we obtain the desired eigensolution $\ket{F({\bf N})}$, the cumulant generating function can be computed,
\begin{align}
Q(\lambda)=-\frac{1}{\delta}\ln\braket{F^{\dagger}({\bf N})|{\rm e}^{\hat{L}_{\lambda}\delta}|F({\bf N})} \text{,}
\end{align}
with $\delta$ being a very small real number.

\section{Numerical Results}\label{sec_results}

\begin{table}
\caption{The values of quantities and parameters specified in numerical computations.}
\begin{center}
\begin{tabular}{>{\centering\arraybackslash}m{8cm}|>{\centering\arraybackslash}m{2cm}}
\hline
\hline
diffusion coefficient, $D$   &  $0.01$  \bigstrut \\ \hline
width of each cell, $\Delta x$   &  $0.1$  \bigstrut \\ \hline
number of cells in discretized channel, $L$   &  $5$  \bigstrut \\ \hline
number of particle in left-reservoir cell, $\bar{N}_{\rm L}$   &  $9$  \bigstrut \\ \hline
number of particle in right-reservoir cell, $\bar{N}_{\rm R}$   &  $3$  \bigstrut \\ \hline
dimension of $a_i^+$, $a_i^-$, and $a_i$ after truncation, $N_{\rm max}$   &  $20$  \bigstrut \\ \hline
\hline
\end{tabular}
\end{center}
\label{tab_numerical_values}
\end{table}

\begin{figure}
\begin{center}
\begin{minipage}[t]{0.5\hsize}
\resizebox{1.0\hsize}{!}{\includegraphics{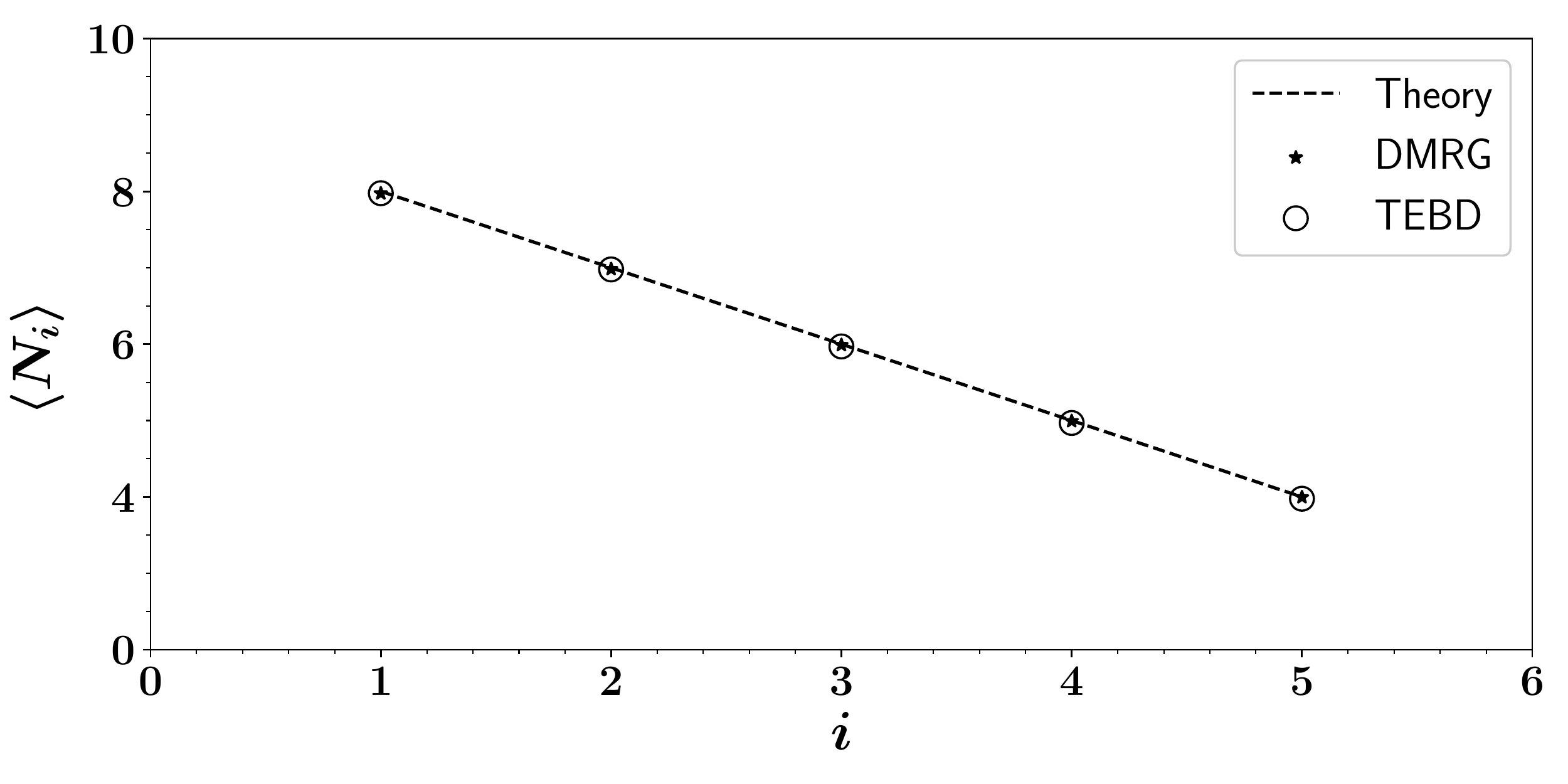}}
\end{minipage}
\end{center}
\caption{Mean particle number $\langle N_i\rangle$. Dash line plots the exact linear profile solution~(\ref{eq_mean}). Asterisks are the numerical results from DMRG approach, while the circles from TEBD approach. The values for numerical computations are listed in Table~\ref{tab_numerical_values}.}
\label{fig_N}
\end{figure}

\par With the tensor-network approaches having already exhibited in the preceding sections, we now dive into the numerical details and present the results. First of all, we define the system with specific numbers which are listed in Table~\ref{tab_numerical_values}. We realize that an arbitrary number of particles can concentrate in a discretized cell of the transport channel. However, the probability of the event that the particle numbers $\{N_i\}$ far exceeds their mean value is exponentially small. This means that we can truncate the state space. The particle numbers in a cell only take the values $N_i=0,1,2,\cdots,N_{\rm max}-1$. Accordingly, the operators $a_i^-$, $a_i^-$, $a$ defined by Eqs.~(\ref{eq_a_minus})-(\ref{eq_a}) have finite dimension $N_{\rm max}\times N_{\rm max}$, as also specified in Table~\ref{tab_numerical_values}. Choosing a proper value of $N_{\rm max}$ is very crucial in numerical computation. Smaller value leads to inaccuracy of the results while larger value costs more computational resources.

\begin{figure}
\begin{center}
\begin{minipage}[t]{0.5\hsize}
\resizebox{1.0\hsize}{!}{\includegraphics{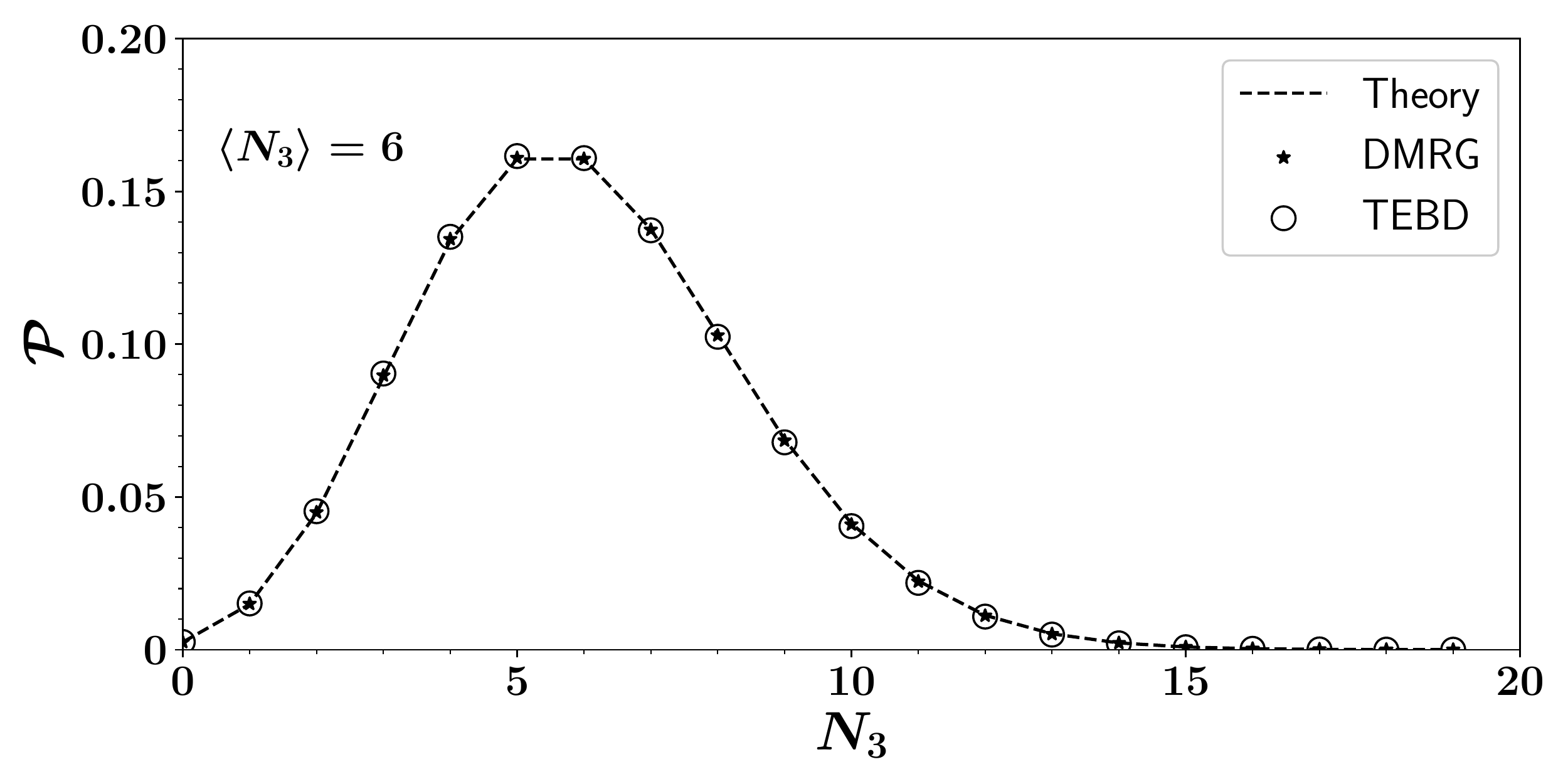}}
\end{minipage}
\end{center}
\caption{Probability distribution of particle number in the third cell ${\cal P}(N_3)$. Dash line plots the exact solution which is a Poisson distribution with the mean $\langle N_3\rangle=6$. The state space for a single cell is 20, stipulated by dimension $N_{\rm max}$ for truncation. Asterisks are the numerical results from DMRG approach, while the circles from TEBD approach. The values for numerical computations are listed in Table~\ref{tab_numerical_values}.}
\label{fig_P}
\end{figure}

\par
\begin{figure}
\begin{center}
\begin{minipage}[t]{0.5\hsize}
\resizebox{1.0\hsize}{!}{\includegraphics{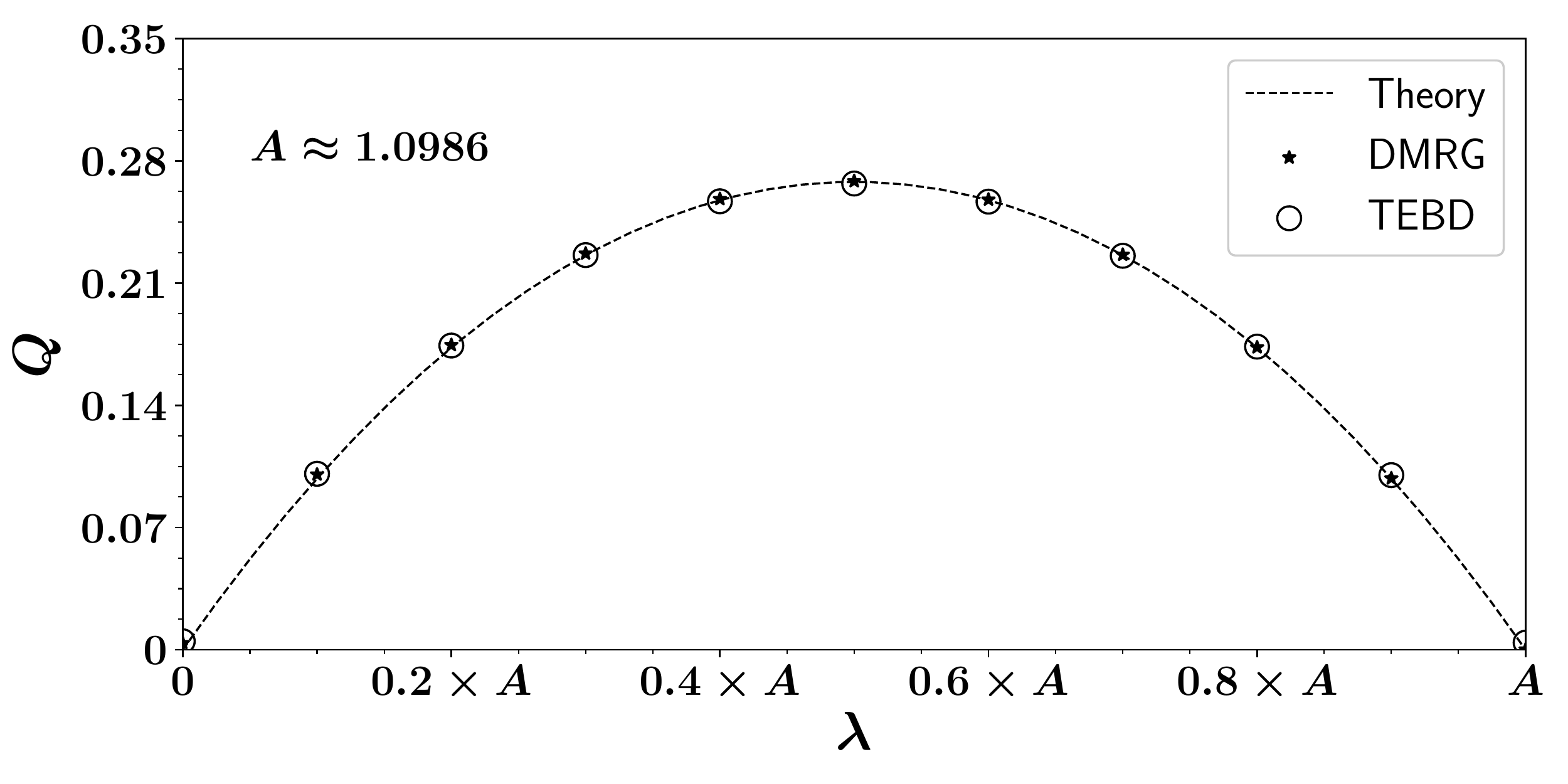}}
\end{minipage}
\end{center}
\caption{Cumulant generating function $Q(\lambda)$. Dash line plots the exact solution~(\ref{eq_CGF}). Asterisks are the numerical results from DMRG approach, while the circles from TEBD approach. The values for numerical computations are listed in Table~\ref{tab_numerical_values}.}
\label{fig_Q}
\end{figure}

 We are now interested in the stationary distribution of the system in steady state. Numerically, this can be obtained with both DMRG and TEBD approaches for the generator $\hat{L}$. The 2-norm probability that the $i$-th cell have $N_i$ particles is calculated as $\tilde{\cal P}(N_i)=\sum_{{\bf N}\backslash N_i}\tilde{\cal P}({\bf N})$. This probability can be obtained by contracting $\ket{\tilde{\cal P}({\bf N})}$ represented as an MPS with an auxiliary MPS given by
\begin{align}
\bra{{\rm auxilary\;MPS}}={\bf 1}_1\otimes\cdots\otimes{\bf 1}_{i-1}\otimes{\bf V}_i\otimes{\bf 1}_{i+1}\otimes\cdots\otimes{\bf 1}_L \text{,}
\end{align}
where ${\bf 1}_j$ is a vector with all elements one, and ${\bf V}_i$ a vector with $N_i$-th element one and others zero. The tensor product $\otimes$ creates a bond of dimension 1, linking the vectors to form an MPS. Then, the 1-norm probability distribution ${\cal P}(N_i)$ can be recovered through Eq.~(\ref{eq_1norm_prob}). The mean value can also be calculated, $\langle N_i\rangle=\sum_{N_i}N_i{\cal P}(N_i)$. Figures~\ref{fig_N} and~\ref{fig_P} respectively plot the profile of mean particle numbers and the probability distribution of particle number in a given cell. The profile is linear, in agreement with the theory (Eq.~\ref{eq_mean}). The probability distribution is Poissonian, also in agreement with the theory. We notice in Figure~\ref{fig_P} that the probability ${\cal P}(N_{\rm max}-1)$ is indeed vanishingly small, manifesting that there is no over-truncation of the state space. Next, we numerically calculate the cumulant generating function $Q(\lambda)$. The counting parameter $\lambda$ takes several discrete values from $0$ to the affinity $A=\ln\left(\bar{N}_{\rm L}/\bar{N}_{\rm R}\right)$. The result is presented in Figure~\ref{fig_Q}. The agreement between the numerical results and the theory (Eq.~\ref{eq_CGF}) is also striking. These agreements strongly manifest validity of both DMRG and TEBD approaches in such an application.

\par Although DMRG and TEBD both give nice results for the system specified with the numbers in Table~\ref{tab_numerical_values}. DMRG is in general much more efficient than TEBD. Because, in the adopted version of DMRG we deals with one tensor at a time, whereas in TEBD we are simultaneously deals two tensors at a time. On the other hand, DMRG also gives more accurate result than TEBD does. Because, in TEBD there is an error associated with the Trotter-Suzuki decomposition.

\par The computer program for numerical computation is coded in C++~\cite{Stroustrup_2013} with the ITensor library~\cite{Fishman_arXiv_2020}. Readers are referred to Ref.~\cite{Psarras_arXiv_2021} where a comprehensive and up-to-date snapshot of software for tensor computations is assembled.

\section{Conclusion and Perspectives}\label{sec_conclusion}

\par In this paper, we studied the diffusive process of Brownian particles in a channel, with a particular interest in the current fluctuation. A nonequilibrium constraint is imposed by placing the channel in between two particle reservoirs with different concentrations. To describe the randomness of the particle transport, the channel is spatially discretized into cells so that a master equation can be written down for the time evolution of the system state specified by the particle numbers in the cells. The counting statistics is performed by modifying the master equation to include a counting parameter for particle transitions between the cells. This gives a tilted generator for the time evolution of the corresponding master equation. The cumulant generating function for the current can be obtained from the leading eigenvalue of the tilted generator. Because the transition rates are linear with the local particle concentration, an analytical solution for the cumulant generating function is obtained, exhibiting the Gallavotti-Cohen symmetry.

\par The focus of this paper is on the application of tensor networks to counting statistics. We presented two approaches, DMRG and TEBD, in details. The tilted generator is written in terms of the local annihilation, creation, particle number operators associated with random particle transitions. The annihilation and creation operators for the reservoir cells are defined differently to make the nonequilibrium constraint sustainable. We also emphasized the differences from their conventional applications for quantum many-body systems. We introduced the so-called 2-norm probability distribution in order to make use of the tensor-network techniques already developed in the quantum-mechanical setting. The conventional probability distribution can be recovered once the corresponding 2-norm probability distribution is obtained. In the DMRG approach, the key point is to construct an MPO for the tilted generator. This is done by constructing two MPOs for different parts of the tilted generator, and then performing tensor summation of these two MPOs. The resulting MPO is of bond dimension 7. We formulated the DMRG approach in such a way that a single tensor is optimized at a time. In the TEBD approach, the tilted evolution operator is approximately decomposed into an ordered product of local ones. Then these local evolution operators are alternatively applied to the an arbitrary distribution represented as an MPS. In both approaches, what we finally obtain is the eigensolution which is then used to calculate the cumulant generating function. When the counting parameter is set to zero, the leading eigensolution of the generator for the master equation represents the stationary distribution for the system in steady state. We compared the profile of mean particle numbers and the probability distribution of particles in a single cell with those given by the analytic solution. We remark here that the joint probability distribution of particles in several cells can also be obtained. We also compared the numerically obtained cumulant generating function with its analytical counterpart. Striking agreements are found for all these comparisons between the numerical results and theory.

\par The tensor-network approaches provides a good means to investigate the fluctuations for a vast majority of realistic systems, e.g., diffusion-reaction systems with many species of particles, systems with long-range electrostatic interactions between particles~\cite{Andrieux_JStatMech_2009, Gu_PhysRevE_2018, Gu_PhysRevE_2019}, or systems of complex geometries with more than two particle reservoirs. Moreover, we notice that the tensor-network approaches, especially DMRG, enable a very accurate determination of the cumulant generating function. This makes it possible to test Onsager reciprocal relations~\cite{Onsager_PhysRev_1931a, Onsager_PhysRev_1931b, Casimir_RevModPhys_1945} and their generalizations to nonlinear regimes~\cite{Andrieux_JChemPhys_2004, Andrieux_JStatMech_2007, Gaspard_NewJPhys_2013, Barbier_JPhysA_2018, Barbier_JPhysA_2019, Barbier_JPhysA_2020} for spatially extended systems. It also makes it possible to study the mathematical properties for statistical cumulants. For example, whether the limited number of cumulants can be used to infer the affinity~\cite{Gu_JStatMech_2020}. To conclude, an accurate determination of the cumulant generating function possibly reveals many underlying physics.

\section*{Acknowledgments}

Financial support from National Science Foundation of China under the Grants No. 11775001, 11825501, 12147162 is acknowledged.

\appendix

\section{Formal Derivation of the Cumulant Generating Function}\label{app_CGF}

\par In this appendix, we provide a formal derivation of the cumulant generating function~(\ref{eq_CGF}) with the method introduced in Ref.~\cite{Gaspard_JStatMech_2018a}. We now consider the probability ${\cal P}(Z,N_1,\cdots,N_L,t)$ that the cells contain given particle numbers and that the signed cumulated number $Z$ of particles is transferred from the $I$-th to the $(I+1)$-th cells during time interval $[0,\,t]$. The time evolution of this probability is ruled by the following master equation,
\begin{align}
\frac{{\rm d}{\cal P}}{{\rm d}t}=& k\bar{N}_{\rm L}\left({\rm e}^{-\partial_{N_1}}-1\right){\cal P}+k\left({\rm e}^{+\partial_{N_1}}-1\right)N_1{\cal P}+k\left({\rm e}^{+\partial_{N_1}}{\rm e}^{-\partial_{N_2}}-1\right)N_1{\cal P} \nonumber \\
& +\sum_{i=2}^{I-1}\left[k\left({\rm e}^{+\partial_{N_i}}{\rm e}^{-\partial_{N_{i-1}}}-1\right)N_i{\cal P}+k\left({\rm e}^{+\partial_{N_i}}{\rm e}^{-\partial_{N_{i+1}}}-1\right)N_i{\cal P}\right] \nonumber \\
& +k\left({\rm e}^{+\partial_{N_I}}{\rm e}^{-\partial_{N_{I-1}}}-1\right)N_I{\cal P}+k\left({\rm e}^{+\partial_{N_I}}{\rm e}^{-\partial_{N_{I+1}}}{\rm e}^{-\partial_Z}-1\right)N_I{\cal P} \nonumber \\
& +k\left({\rm e}^{+\partial_{N_{I+1}}}{\rm e}^{-\partial_{N_I}}{\rm e}^{+\partial_Z}-1\right)N_{I+1}{\cal P}+k\left({\rm e}^{+\partial_{N_{I+1}}}{\rm e}^{-\partial_{N_{I+2}}}-1\right)N_{I+1}{\cal P} \nonumber \\
& +\sum_{i=I+2}^{L-1}\left[k\left({\rm e}^{+\partial_{N_i}}{\rm e}^{-\partial_{N_{i-1}}}-1\right)N_i{\cal P}+k\left({\rm e}^{+\partial_{N_i}}{\rm e}^{-\partial_{N_{i+1}}}-1\right)N_i{\cal P}\right] \nonumber \\
& +k\bar{N}_{\rm R}\left({\rm e}^{-\partial_{N_L}}-1\right){\cal P}+k\left({\rm e}^{+\partial_{N_L}}{\rm e}^{-\partial_{N_{L-1}}}-1\right)N_L{\cal P}+k\left({\rm e}^{+\partial_{N_L}}-1\right)N_L{\cal P} \text{.} \label{eq_extended_master_equation}
\end{align}
We then define the moment generating function
\begin{align}
G(\eta,s_1,\cdots,s_L,t)=\sum_{Z,N_1,\cdots,N_L}\eta^Z\left(\prod_is_i^{N_i}\right){\cal P}(Z,N_1,\cdots,N_L,t) \text{.}
\end{align}
where $\eta\equiv{\rm e}^{-\lambda}$ with $\lambda$ being the counting parameter for the particle transfers $Z$. From the master equation~(\ref{eq_extended_master_equation}), we can easily obtain the following first-order partial differential equation,
\begin{align}
\partial_tG & + \left[k(s_1-1)+k(s_1-s_2)\right]\partial_{s_1}G \nonumber \\
& +\sum_{i=2}^{I-1}\left[k(s_i-s_{i-1})+k(s_i-s_{i+1})\right]\partial_{s_i}G \nonumber \\
& +\left[k(s_I-s_{I-1})+k(s_I-\eta s_{I+1})\right]\partial_{s_I}G \nonumber \\
& +\left[k(s_{I+1}-\eta^{-1}s_I)+k(s_{I+1}-s_{I+2})\right]\partial_{s_{I+1}}G \nonumber \\
& +\sum_{i=I+1}^{L-1}\left[k(s_i-s_{i-1})+k(s_i-s_{i+1})\right]\partial_{s_i}G \nonumber \\
& + \left[k(s_L-s_{L-1})+k(s_L-1)\right]\partial_{s_L}G \nonumber \\
& =\left[k\bar{N}_{\rm L}(s_1-1)+k\bar{N}_{\rm R}(s_L-1)\right]G \text{,}
\end{align}
which, in vectorial notations, can be written in the following form,
\begin{align}
\partial_tG+({\boldsymbol{\sf L}}\cdot{\bf s}+{\bf f})\cdot\partial_{\bf s}G=({\bf g}\cdot{\bf s}+h)G \label{eq_differential_equation}
\end{align}
where
\begin{align}
{\boldsymbol{\sf L}}\equiv
k\begin{pmatrix}
2 & -1 & & & & & & \\
-1 & 2 & -1 & & & & & \\
& \ddots & \ddots & \ddots & & & & \\
& & -1 & 2 & -\eta & & & \\
& & & -\eta^{-1} & 2 & -1 & & \\
& & & & \ddots & \ddots & \ddots & \\
& & & & & -1 & 2 & -1 \\
& & & & & & -1 & 2
\end{pmatrix} \text{,}
\end{align}
\begin{align}
{\bf s}\equiv
\begin{pmatrix}
s_1 \\
s_2 \\
\vdots \\
s_{L-1} \\
s_L
\end{pmatrix}\text{,}
\hspace{1cm}
{\bf f}\equiv-
\begin{pmatrix}
k \\
0 \\
\vdots \\
0 \\
k
\end{pmatrix}\text{,}
\hspace{2cm}
{\bf g}\equiv
\begin{pmatrix}
k\bar{N}_{\rm L} \\
0 \\
\vdots \\
0 \\
k\bar{N}_{\rm R}
\end{pmatrix}\text{,}
\end{align}
and
\begin{align}
h\equiv-k\bar{N}_{\rm L}-k\bar{N}_{\rm R} \text{.}
\end{align}
By setting $\eta=1$ in the matrix ${\boldsymbol{\sf L}}$, we can define ${\boldsymbol{\sf L}}_0$. So, we have the relations
\begin{align}
& {\bf f}=-{\boldsymbol{\sf L}}_0\cdot{\bf 1} \text{,} \\
& h=-{\bf g}\cdot{\bf 1} \text{,}
\end{align}
where ${\bf 1}$ denotes the vector with all elements being equal to one. Besides, the stationary values of mean particle numbers are given by
\begin{align}
{\bf\Gamma}_0={\boldsymbol{\sf L}}_0^{-1{\rm T}}\cdot{\bf g} \text{.} \label{eq_Gamma_0}
\end{align}
 Here, ${\rm T}$ denotes the transpose. The first-order partial differential equation~(\ref{eq_differential_equation}) can be solved by the method of characteristics~\cite{Gardiner_2004}. The solution is given by
\begin{align}
G=G_0\exp\left[{\bf g}\cdot{\boldsymbol{\sf L}}^{-1}\cdot\left({\boldsymbol{\sf I}}-{\rm e}^{-{\boldsymbol{\sf L}}t}\right)\cdot\left({\bf s}+{\boldsymbol{\sf L}}^{-1}\cdot{\bf f}\right)+\left(h-{\bf g}\cdot{\boldsymbol{\sf L}}^{-1}\cdot{\bf f}\right)t\right] \text{,}
\end{align}
where $G_0$ denotes the initial condition, and ${\boldsymbol{\sf I}}$ the identity matrix. The cumulant generating function for the current can be obtained,
\begin{align}
Q(\lambda)\equiv\lim_{t\to\infty}-\frac{1}{t}\ln\left[G(\eta={\rm e}^{-\lambda},{\bf 1},t)\right]={\bf g}\cdot\left({\bf 1}+{\boldsymbol{\sf L}}^{-1}\cdot{\bf f}\right) \text{.} \label{eq_Q_1}
\end{align}
We observe that
\begin{align}
{\boldsymbol{\sf L}}=\boldsymbol{\sf M}\cdot{\boldsymbol{\sf L}}_0\cdot\boldsymbol{\sf M}^{-1} \text{,}
\end{align}
where
\begin{align}
\boldsymbol{\sf M}=\eta\boldsymbol{\sf P}_{\rm L}+\boldsymbol{\sf P}_{\rm R}
\end{align}
with the projection matrices
\begin{align}
\boldsymbol{\sf P}_{\rm L}=
\begin{pmatrix}
1 & \cdots & 0 & 0 & \cdots & 0 \\
\vdots & \ddots & \vdots & \vdots & \ddots & \vdots \\
0 & \cdots & 1 & 0 & \cdots & 0 \\
0 & \cdots & 0 & 0 & \cdots & 0 \\
\vdots & \ddots & \vdots & \vdots & \ddots & \vdots \\
0 & \cdots & 0 & 0 & \cdots & 0 \\
\end{pmatrix}
\hspace{1cm}\text{and}\hspace{1cm}
\boldsymbol{\sf P}_{\rm R}=
\begin{pmatrix}
0 & \cdots & 0 & 0 & \cdots & 0 \\
\vdots & \ddots & \vdots & \vdots & \ddots & \vdots \\
0 & \cdots & 0 & 0 & \cdots & 0 \\
0 & \cdots & 0 & 1 & \cdots & 0 \\
\vdots & \ddots & \vdots & \vdots & \ddots & \vdots \\
0 & \cdots & 0 & 0 & \cdots & 1 \\
\end{pmatrix} \text{.}
\end{align}
The identity matrix in $\boldsymbol{\sf P}_{\rm L}$ is of dimension $I\times I$, while the identity matrix in $\boldsymbol{\sf P}_{\rm R}$ is $(L-I)\times(L-I)$. Since the projection matrices satisfy the condition $\boldsymbol{\sf P}_{\rm L}+\boldsymbol{\sf P}_{\rm R}={\boldsymbol{\sf I}}$, we thus have
\begin{align}
& \boldsymbol{\sf M}={\boldsymbol{\sf I}}+(\eta-1)\boldsymbol{\sf P}_{\rm L} \text{,} \\
& \boldsymbol{\sf M}^{-1}={\boldsymbol{\sf I}}+(\eta^{-1}-1)\boldsymbol{\sf P}_{\rm L} \text{.}
\end{align}
From the above related expressions, the cumulant generating function~(\ref{eq_Q_1}) can be written in the following form,
\begin{align}
Q(\lambda) & ={\bf g}\cdot\left({\boldsymbol{\sf I}}-\boldsymbol{\sf M}\cdot{\boldsymbol{\sf L}}_0^{-1}\cdot\boldsymbol{\sf M}^{-1}\cdot{\boldsymbol{\sf L}}_0 \right)=W_+\left(1-{\rm e}^{-\lambda}\right)+W-\left(1-{\rm e}^{+\lambda}\right) \text{,} \label{eq_Q_2}
\end{align}
with the global transition rates given by
\begin{align}
& W_+={\bf\Gamma}_0\cdot{\boldsymbol{\sf L}}_0\cdot\boldsymbol{\sf P}_{\rm L}\cdot{\boldsymbol{\sf L}}_0^{-1}\cdot\boldsymbol{\sf P}_{\rm R}\cdot{\boldsymbol{\sf L}}_0\cdot{\bf 1} \text{,} \\
& W_-={\bf\Gamma}_0\cdot{\boldsymbol{\sf L}}_0\cdot\boldsymbol{\sf P}_{\rm R}\cdot{\boldsymbol{\sf L}}_0^{-1}\cdot\boldsymbol{\sf P}_{\rm L}\cdot{\boldsymbol{\sf L}}_0\cdot{\bf 1} \text{.}
\end{align}
These two global transition rates can be further developed as
\begin{align}
W_+=k^2\bar{N}_{\rm L}\left({\boldsymbol{\sf L}}_0^{-1}\right)_{1L} \hspace{1cm}\text{and}\hspace{1cm} W_-=k^2\bar{N}_{\rm R}\left({\boldsymbol{\sf L}}_0^{-1}\right)_{L1} \text{.}
\end{align}
Inverting the matrix ${\boldsymbol{\sf L}}_0$, we get
\begin{align}
\left({\boldsymbol{\sf L}}_0^{-1}\right)_{ij}=
\begin{cases}
\frac{i}{k(L+1)}(L+1-j) & \text{if}  \quad i\leq j \text{,} \\
& \vspace{-0.3cm} \\
\frac{j}{k(L+1)}(L+1-i) & \text{if}  \quad i>j \text{.}
\end{cases}
\end{align}
So, the two global transition rates are calculated as
\begin{align}
W_+=\frac{k\bar{N}_{\rm L}}{L+1} \hspace{1cm}\text{and}\hspace{1cm} W_-=\frac{k\bar{N}_{\rm R}}{L+1} \text{,}
\end{align}
finally giving
\begin{align}
Q(\lambda)=\frac{k\bar{N}_{\rm L}}{L+1}\left(1-{\rm e}^{-\lambda}\right)+   \frac{k\bar{N}_{\rm R}}{L+1}\left(1-{\rm e}^{\lambda}\right) \text{.}
\end{align}

\printbibliography[title={References}]

\end{document}